\def\Box{\kern1pt\vbox{\hrule height 1.2pt\hbox{\vrule width 1.2pt\hskip 3pt
   \vbox{\vskip 6pt}\hskip 3pt\vrule width 0.6pt}\hrule height 0.6pt}\kern1pt}
\def\gtwid{\mathrel{\raise.3ex\hbox{$>$\kern-.75em\lower1ex\hbox{$\sim$}}}}
\def\ltwid{\mathrel{\raise.3ex\hbox{$<$\kern-.75em\lower1ex\hbox{$\sim$}}}}

\documentclass[12pt]{article}
\usepackage{amsmath}
\newcommand{\be}{\begin{equation}}
\newcommand{\ee}{\end{equation}}

\begin{document}
\begin{titlepage}
\begin{flushright}
UFIFT-HEP-03-02 \\ astro-ph/0302030
\end{flushright}
\vspace{.4cm}
\begin{center}
\textbf{A Nonlocal Metric Formulation of MOND}
\end{center}
\begin{center}
M. E. Soussa$^{\dagger}$ and R. P. Woodard$^{\ddagger}$
\end{center}
\begin{center}
\textit{Department of Physics \\ University of Florida \\
Gainesville, FL 32611 USA}
\end{center}

\begin{center}
ABSTRACT
\end{center}
We study a class of nonlocal, but causal, covariant and conserved field 
equations for the metric. Although nonlocal, these equations do not seem to
possess extra graviton solutions in weak field perturbation theory. Indeed,
the equations reduce to those of general relativity when the Ricci scalar 
vanishes throughout spacetime. When a static matter source is present we show
how these equations can be adjusted to reproduce Milgrom's Modified Newtonian 
Dynamics in the weak field regime, while reducing to general relativity for 
strong fields. We compute the angular deflection of light in the weak field 
regime and demonstrate that it is the same as for general relativity, 
resulting in far too little lensing if no dark matter is present. We also 
study the field equations for a general Robertson-Walker geometry. An
interesting feature of our equations is that they become conformally invariant
in the MOND limit.

\begin{flushleft}
\end{flushleft}
\vspace{.4cm}
\begin{flushleft}
$^{\dagger}$ e-mail: soussam@phys.ufl.edu \\
$^{\ddagger}$ e-mail: woodard@phys.ufl.edu
\end{flushleft}
\end{titlepage}

\section{Introduction}

Milgrom's Modified Newtonian Dynamics (MOND) \cite{Milgrom1} is an empirical 
alternative to invoking dark matter as an explanation for the motions of 
cosmological systems which are subject to very small accelerations. In one
formulation the force $\vec{F}_{\rm Newt}$ on a point mass $m$ is given by 
Newtonian gravity, but the second law becomes nonlinear in the acceleration 
$\vec{a}$,
\begin{equation}
\vec{F}_{\rm Newt} = m \mu\Big(\frac{a}{a_0}\Bigr) \vec{a} \qquad {\rm where} 
\qquad \mu(x) = \begin{cases}
1 & \text{$\forall x \gg 1$} \; , \\
x & \text{$\forall x \ll 1$} \; . \end{cases}
\end{equation}
The numerical value of $a_0$ has been determined by fitting to the rotation
curves of nine well-measured galaxies \cite{BBS},
\begin{equation}
a_0 = (1.20 \pm .27) \times 10^{-10}~{\rm m/s}^2 \; .
\end{equation}
Because the observed stars and gas contain constituents subject to much 
larger internal accelerations\footnote{For example, were we to think of 
neutral Hydrogen as a classical planetary system, the proton would be 
subject to an acceleration of about $10^{19}~{\rm m/s}^2$ about the atomic
barycenter!} it is preferable to view MOND as a modification of the 
gravitational force at low accelerations. That is, the second law takes its
traditional form $\vec{F}_{\rm MOND} = m \vec{a}$, but the actual force is 
the following nonlinear function of the force predicted by Newtonian gravity,
\begin{equation}
\vec{F}_{\rm MOND} = f\Big(\frac{F_{\rm Newt}}{m a_0}\Bigr) \vec{F}_{\rm Newt}
\qquad {\rm where} \qquad f(x) = 
\begin{cases}
1 & \text{$\forall x \gg 1$} \; , \\
x^{-\frac12} & \text{$\forall x \ll 1$} \; . \end{cases}
\label{modgrav}
\end{equation}

One consequence of MOND is that the velocities of particles in progressively 
more distant circular orbits approach a constant $v_{\infty}$ which depends
only upon a galaxy's mass \cite{Milgrom2}. For large radius $r$ the Newtonian 
force due to a galaxy of mass $M$ is $F_{\rm Newt} = GMm/r^2$. Since 
$F_{\rm Newt}/m$ must eventually fall below $a_0$, relation (\ref{modgrav}) 
implies that the asymptotic force law is $F_{\rm MOND} = m \sqrt{a_0 G M}/r$. 
Applying the second law gives,
\begin{equation}
\frac{\sqrt{a_0 G M}}{r} = \frac{v_{\infty}^2}{r} \qquad \Longrightarrow
\qquad a_0 G M = v_{\infty}^4 \; .
\end{equation}
With no dark matter a galaxy's luminosity $L$ should be a constant times its
mass, although the constant will depend upon the type of galaxy. Thus one 
expects $L \sim v_{\infty}^4$, which is the observed Tully-Fisher relation 
\cite{TF}.

MOND provides an excellent fit to the rotation curves of all known types of 
galaxies, using only the measured distributions of gas and stars and fitted 
mass-to-luminosity ratios for gas and stars. The recent review by Sanders and 
McGaugh \cite{SanMc} summarizes the increasingly compelling data and lists 
the primary sources. It is significant that the fitted mass-to-luminosity 
ratios are not unreasonable \cite{BdJ}. Especially significant is that MOND 
agrees in detail even with low surface brightness galaxies \cite{SMEB,EBSM}, 
objects for which the MOND regime ($a \ltwid a_0$) applies throughout and for 
which no detailed measurements had been made when MOND was proposed. On the 
other hand, some dark matter must be invoked to explain the temperature and 
density profiles of large galaxy clusters \cite{ASQ}. (It it has been 
suggested that $eV$-mass neutrinos might provide this without affecting the 
galactic results \cite{Sanders1}.) And a very recent analysis of data from the 
Sloan Digital Sky Survey asserts that satellites of isolated galaxies violate 
MOND when care is taken to exclude interlopers \cite{Petal}.

The successes of MOND might be a numerical coincidence \cite{KapTur}. They 
may result instead from galaxy formation and evolution flowing towards some 
yet-unrecognized attractor solution through conventional physics. Or they 
might signal a real modification of gravity in the regime of very low 
acceleration. Although the issue cannot be decided at this time, the number 
and quality of relevant observations which are technically feasible 
\cite{SelKos} indicate that it may be resolved in the near future. If MOND 
could be convincingly embedded in a larger, metric formulation of gravity, 
that theory could {\it already} be tested against a wide variety of data from 
lensing \cite{MorTurn1,MorTurn2}, cosmology \cite{Felten}, and structure 
formation \cite{Sanders2}. As it is, one must make assumptions, whereupon 
comparison with observation is as much a test of these assumptions as it is 
of MOND. It is therefore opportune to consider what sort of fundamental theory 
might reduce to MOND in the appropriate limit.

There is a satisfactory nonrelativistic potential formulation for the relation 
between the MOND force $\vec{F}_{\rm MOND} = -\vec{\nabla} \phi$ and the 
mass density $\rho_m$. It is given by the Lagrangian of Bekenstein and 
Milgrom \cite{BekMil},
\begin{equation}
{\cal L} = - \rho_m \phi - \frac{a_0^2}{8 \pi G} F\Bigl(\frac{\Vert \vec{\nabla}
\phi \Vert^2}{a_0^2}\Bigr) \qquad {\rm where} \qquad \mu(x) = F'(x^2) \; .
\end{equation}
This is very important because it establishes that MOND conserves energy,
momentum and angular momentum. However, it does not extend the theory 
sufficiently to test lensing and cosmology. A generally coordinate invariant,
scalar-metric extension exists but it contains dynamical scalar degrees of
freedom which can propagate acausally \cite{BekMil}.

We wish here to consider a different possibility. Suppose that general 
relativity really is the fundamental theory of gravity, but that its effective
action contains large quantum corrections from infrared virtual particles. 
Weinberg showed that infrared effects in quantum gravity are no stronger than 
those of QED for the case of zero cosmological constant \cite{Wein1}. However, 
a nonzero cosmological constant would preserve the graviton's masslessness 
while subjecting it to interactions of canonical dimension three. Recall that 
infrared effects become stronger as massless particles are coupled with lower 
dimension interactions, and that they are already nonperturbatively strong for 
massless gluons with the dimension four coupling of QCD. 

These considerations are the basis for a daring proposal to simultaneously 
resolve the (old) cosmological constant problem \cite{Wein2,Carroll} and 
provide a natural model of inflation in which scalars play no part. The idea
\cite{TsWo1} is that the bare cosmological constant is actually GUT-scale, 
which leads to an initial period of inflation during the early universe. What
brings inflation to an end is the gravitational attraction between the 
ever-increasing numbers of infrared gravitons ripped out of the vacuum by the 
rapid expansion. One can follow this process as long as the slowing is weak,
and explicit computations confirm that the effect must eventually become 
nonperturbatively strong \cite{TsWo2}. 

If this proposal is correct, the post-inflationary universe would be described 
by the nonlocal effective action which prevails after the breakdown of 
perturbation theory. We cannot compute reliably in this regime, but nothing 
prevents one from making guesses about the form of this effective action. The
simplest class of guesses which give a plausible end for inflation involve
acting the inverse covariant d'Alembertian on the Ricci scalar \cite{TsWo3}.
We will refer to this as {\it the small potential},
\begin{equation}
\varphi[g] \equiv \frac{1}{\Box} R \qquad {\rm where} \qquad \Box \equiv
\frac{1}{\sqrt{-g}} \partial_{\mu} \Bigl( \sqrt{-g} g^{\mu\nu} \partial_{\nu}
\Bigr) \; . \label{smallpot}
\end{equation}
(We use a spacelike metric with $R_{\mu\nu} \equiv \Gamma^{\rho}_{~\nu\mu , 
\rho} - \Gamma^{\rho}_{~ \rho \mu , \nu} + \Gamma^{\rho}_{~ \rho \sigma} 
\Gamma^{\sigma}_{~ \nu \mu} - \Gamma^{\rho}_{~ \nu \sigma} \Gamma^{\sigma}_{~
\rho \mu}$.) This paper will not consider inflation or the cosmological 
constant. We shall rather explore embedding MOND in a nonlocal Lagrangian 
of the form,
\begin{equation}
{\cal L} = \frac{c^4}{16 \pi G} \Bigl[ R + c^{-4} a_0^2 {\cal F}\Bigl(c^4
a_0^{-2} g^{\mu \nu} \varphi_{,\mu} \varphi_{,\nu}\Bigr) \Bigr] \sqrt{-g} \; .
\label{genmod}
\end{equation}

A desirable feature of this class of models is that it involves only the 
metric. Although the field equations are nonlocal, they do not seem to possess
additional graviton solutions in weak field perturbation theory. To see this,
expand the metric about flat space as usual,
\begin{equation}
g_{\mu\nu} = \eta_{\mu\nu} + h_{\mu\nu} \; .
\end{equation}
It is an elementary exercise to show that the Ricci scalar is,
\begin{equation}
R = h^{\mu\nu}_{~~ , \mu\nu} - h^{\mu~,\nu}_{~\mu~~\nu} + O(h^2) \; ,
\end{equation}
where graviton indices are raised and lowered using the Lorentz metric as 
usual. Now impose de Donder gauge,
\begin{equation}
h^{\mu}_{~\nu , \mu} - \frac12 h^{\mu}_{~\mu , \nu} = 0 \; ,
\end{equation}
to show that the small potential is local in the weak field limit,
\begin{equation}
\varphi[\eta + h] = -\frac12 h^{\mu}_{~\mu} + O(h^2) \; .
\end{equation}
Since the Lagrangian depends upon the first derivative of $\varphi$, this
theory cannot acquire higher derivative solutions in weak field perturbation
theory. Indeed, all solutions to the source-free Einstein equations must
be solutions to this theory because they have $R = 0$ throughout spacetime,
which implies $\varphi = 0$ as well. The correction term in (\ref{genmod})
changes only gravity's response to sources, without adding new weak field
dynamical degrees of freedom. So these models seem worthy of closer study.

In section 2 we derive the form a static and spherically symmetric metric 
must take to reproduce galactic rotation curves in the MOND regime. We also 
work out the deflection of light in the weak field limit. In section 3 we
employ a formal shortcut to obtain the causal field equations which would 
come from a Schwinger-Keldysh effective action \cite{Jordan}. We also 
explicitly demonstrate conservation. The nonlocal field equations are 
specialized to a static, spherically symmetric geometry in section 4. We 
first prove, for general interpolating function ${\cal F}(x)$, that any MOND 
terms must drop out of the formula for the deflection of light. This is the 
decisive point of our analysis because the existing lensing data can only be 
made consistent with the absence of dark matter if MOND terms contribute 
\cite{MorTurn2}. No member of this class of models can therefore represent a 
phenomenologically viable extension of MOND. However, it is possible to 
choose the interpolating function ${\cal F}(x)$ to reproduce MOND, and we 
derive the required form. Section 5 explores the model's impact on cosmology, 
more  to illustrate what can be done with a complete theory than because this 
model is viable. In section 6 we discuss more general models which might 
avoid the lensing disaster.

\section{Phenomenological constraints}

MOND was developed as an alternative to halos of dark matter surrounding 
nearly static, cosmological sources. Far from such a system one would expect 
the asymptotic geometry to be spherically symmetric and static:
\begin{equation}
ds^2 \equiv g_{\mu\nu}(x) dx^{\mu} dx^{\nu} = -B(r) dt^2 + A(r) dr^2 + r^2
d\Omega^2 \; . \label{sphrm}
\end{equation}
We first show that rotation curves fix $B(r)$ in the MOND limit but tell
us nothing about $A(r)$. We then discuss the magnitude of the effect. The
section closes by working out the angular deflection of light for a general
class of weak field limits.

Let $\chi^{\mu}(t)$ be the worldline of a test particle moving in the 
geometry described by (\ref{sphrm}). If only the gravitational force is
significant the particle's worldline obeys the geodesic equation:
\begin{equation}
\ddot{\chi}^{\mu}(t) + \Gamma^{\mu}_{~\rho\sigma}\!\Bigl(\chi(t)\Bigr) 
\dot{\chi}^{\rho}(t) \dot{\chi}^{\sigma}(t) = 0 \; . \label{geo}
\end{equation}
The nonzero connection components derived from (\ref{sphrm}) are,
\begin{eqnarray}
&& \Gamma^t_{~tr} = \frac{B'}{2 B} \; , \; \Gamma^r_{~ tt} = \frac{B'}{2 A} 
\; , \; \Gamma^r_{~rr} = \frac{A'}{2 A} \; , \; \Gamma^r_{~ \theta \theta} 
= -\frac{r}{A} \; , \; \Gamma^r_{~ \phi \phi} = -\frac{r}{A} \sin^2(\theta) 
\; , \nonumber \\
&& \Gamma^{\theta}_{~ \theta r} = \frac{1}{r} \; , \; \Gamma^{\phi}_{~ \phi r} 
= \frac{1}{r} \; , \; \Gamma^{\theta}_{~ \phi \phi} = -\sin(\theta) \cos(
\theta) \; , \; \Gamma^{\phi}_{~ \phi \theta} = \cot(\theta) \; . \label{Gam}
\end{eqnarray}
Now specialize to the case of circular motion,
\begin{equation}
\Bigl(\chi^t,\chi^r,\chi^\theta,\chi^\phi\Bigr) = \Bigl(ct,r,\frac{\pi}2,
\phi(t)\Bigr) \; .
\end{equation}
With our connection (\ref{Gam}) the $\mu=t$ and $\mu=\theta$ components 
of the geodesic equation are tautologies. The $\mu=\phi$ component just 
says $\dot{\phi}$ is constant. Only the $\mu=r$ component of the geodesic
equation is nontrivial,
\begin{equation}
\frac{B'}{2 A} - \frac{r}{A} \frac{\dot{\phi}^2}{c^2} = 0 \; .
\end{equation}

Note that $A(r)$ factors out! For circular orbits the velocity has the 
Euclidean relation to the angular velocity, $v = r \dot{\phi}$. In the MOND 
limit $v^2$ approaches the constant $v^2_{\infty} = \sqrt{a_0 G M}$, so the 
MOND limit for $B(r)$ must obey,
\begin{equation}
B'(r) \longrightarrow \frac{2}{r} \sqrt{\frac{a_0 G M}{c^4}} \; .
\end{equation}

It is well at this point to consider the size of things. A large galaxy 
might have a mass in stars and gas of $M \sim 10^{11} \times M_{\odot} 
\sim 10^{41}~{\rm kg}$. Such a galaxy would enter the MOND regime at a 
radius of about,
\begin{equation}
R_{\rm gal} \sim \sqrt{\frac{G M}{a_0}} \sim 10^{20}~{\rm m} \; .
\end{equation}
For weak fields we can write,
\begin{equation}
A(r) = 1 + a(r) \qquad , \qquad B(r) = 1 + b(r) \; ,
\end{equation}
where $\vert a(r) \vert \ll 1$ and $\vert b(r) \vert \ll 1$. A useful
phenomenological ansatz for the asymptotic behavior of the weak fields is,
\begin{equation}
a(r) \longrightarrow \delta_1 \frac{G M}{c^2 r} + \epsilon_1 
\sqrt{\frac{a_0 G M}{c^4}} \; , \; b(r) \longrightarrow \delta_2 
\frac{G M}{c^2 r} + \epsilon_2 \sqrt{\frac{a_0 G M}{c^4}} \ln\Bigl(
\frac{r}{R_{\rm gal}}\Bigr) \; . 
\label{ansatz}
\end{equation}
We have just seen that MOND predicts $\epsilon_2 = 2$ but says nothing about
the other parameters. With just the isolated galaxy, general relativity gives
$\delta_1 = -\delta_2 = 2$ and $\epsilon_1 = \epsilon_2 = 0$. If an isothermal 
halo of dark matter is added, whose density is chosen to reproduce 
$v^2_{\infty} = \sqrt{a_0 G M}$, general relativity gives $\delta_1 = 
-\delta_2 = \epsilon_1 = \epsilon_2 = 2$.

One might worry that the logarithmic growth of $b(r)$ in (\ref{ansatz}) must 
eventually invalidate the weak field approximation but this is not a practical 
concern. For the large galaxy considered previously the small parameters
multiplying the $\delta$'s and $\epsilon$'s are,
\begin{equation}
\frac{G M}{c^2 r} \sim \frac{R_{\rm gal}}{r} \times 10^{-6} \qquad , \qquad 
\sqrt{\frac{a_0 G M}{c^4}} \sim 10^{-6} \; .
\end{equation}
The difference $b(r)$ from the onset of MOND all the way to the current 
horizon ($R_{\rm hor} \sim 10^{26}~{\rm m}$) is,
\begin{equation}
b(R_{\rm hor}) - b(R_{\rm gal}) \sim -\delta_2 \times 10^{-6} + \epsilon_2
\times 10^{-5} \; .
\end{equation}
The weak field regime is therefore applicable throughout the Hubble volume.

Another numerical fact worthing noting is that the natural length associated
with the MOND acceleration $a_0$ is larger than the Hubble radius,
\begin{equation}
\frac{c^2}{a_0} \sim 10^{27}~{\rm m} \; .
\end{equation}
This has the important consequence that $r a_0/c^2$ is negligible on galaxy
and galactic cluster scales, so that powers of $r$ do not necessarily 
distinguish ``large'' and ``small'' terms. Consider, for example, two possible
contributions to the weak fields,
\begin{equation}
\frac{a_0 G M}{c^4} \ln\left(\frac{r}{R_{\rm gal}}\right) \ll 
\frac{G M}{c^2 r} \; .
\end{equation}
The left hand term is negligible with respect to the right hand side, even 
though the latter falls off with $r$ whereas the former actually grows.

We can now consider what MOND says about lensing. The angular deflection of 
light can be expressed in terms of the turning point $R_0$,
\begin{equation}
{\Delta \phi} = 2 \int_{R_0}^{\infty} \frac{dr}{r} \frac{A^{\frac12}(r)}{
\sqrt{\left(\frac{r}{R_0}\right)^2 \frac{B(R_0)}{B(r)} - 1}} - \pi \; .
\end{equation}
This can be expanded in powers of the weak fields and then simplified using
the change of variables $r = R_0 \sec(\theta)$,
\begin{alignat}{1}
{\Delta \phi} & = 2 \int_{R_0}^{\infty} \frac{dr}{r} \frac{1}{\sqrt{
\left(\frac{r}{R_0}\right)^2 - 1}} \left\{1 + \frac{a(r)}{2} - \frac12 
\frac{b(R_0) - b(r)}{1 - \left(\frac{R_0}{r}\right)^2} + \dots\right\}-\pi ,\\
& = \!\! \int_0^{\frac{\pi}2} \!\! d\theta \left\{ a\Bigl( R_0 \sec(\theta) 
\Bigr) - \csc^2(\theta) \Bigl[b\Bigl(R_0\Bigr) - b\Bigl(R_0 \sec(\theta) 
\Bigr)\Bigr] + \dots \right\} \! .
\end{alignat}

Substituting the general ansatz (\ref{ansatz}) gives,
\begin{equation}
{\Delta \phi} = \Bigl(\delta_1 - \delta_2\Bigr) \frac{G M}{c^2 R_0} + 
\Bigl(\epsilon_1 + \epsilon_2\Bigr) \frac{\pi}2 \sqrt{\frac{a_0 G M}{c^4}}
+ \dots \; .
\end{equation}
Without dark matter, general relativity ($\delta_1 = -\delta_2 = 2$ and
$\epsilon_1 = \epsilon_2 = 0$) gives too little deflection at large $R_0$ to 
be consistent with the frequency of lensing by galaxies. General relativity
with an isothermal halo of dark matter ($\delta_1 = - \delta_2 = \epsilon_1 =
\epsilon_2 = 2$) is consistent with the existing data \cite{MorTurn2}. 
For MOND to be similarly consistent with the data requires the sum
$(\epsilon_1 + \epsilon_2)$ to be positive and of order one.

\section{The nonlocal field equations}

The ostensible purpose of this section is to derive the field equations 
associated with (\ref{genmod}). We are not going to quite do that for the
very good reason that one does not get causal field equations by varying
a temporally nonlocal action. To see this, consider the quadratic part of
the effective action for a real scalar field $\phi(x)$ in flat spacetime,
\begin{equation}
\Gamma[\phi] = \frac12 \int d^4y \phi(y) [\partial^2 - m^2] \phi(y) - 
\frac12 \int d^4y \int d^4z \phi(y) M^2(x;y) \phi(z) \; .
\end{equation}
Here $M^2(y;z)$ is the scalar self-mass-squared, which is symmetric under
interchange of $y^{\mu}$ and $z^{\mu}$. Taking the variational derivative 
and ignoring surface terms gives,
\begin{equation}
\frac{\delta \Gamma[\phi]}{\delta \phi(x)} = [\partial^2 - m^2] \phi(x)
- \int d^4x' M^2(x;x') \phi(x') \; .
\end{equation}
If $M^2(x;x') \neq 0$ for some $x^{\prime \mu} = x^{\mu} - \Delta^{\mu}$ in 
the past of $x^{\mu}$ then it is also nonzero for $x^{\prime \mu} = x^{\mu} +
\Delta^{\mu}$ in the future of $x^{\mu}$. Another embarrassing property of 
the conventional effective action field equations is that their solutions 
are not typically real even if the associated field operators are Hermitian. 
This is because the effective action field equations are obeyed by in-out 
matrix elements of the field operators, and the ``in'' state may not evolve 
into the same ``out'' state. 

Both causality and reality can be enforced by employing the field equations 
of the Schwinger-Keldysh effective action \cite{Jordan}. These are obeyed by 
the expectation values of the field operators in the presence of the same 
state, so Hermiticity of the operator implies reality of the solution. In
the Schwinger-Keldysh formalism there are two background fields: $\phi_+(x)$
for the evolution forward from the initial state and $\phi_-(x)$ for the
reverse evolution back to it. The quadratic part of the scalar action has 
the form,
\begin{alignat}{1}
\Gamma[\phi_+;\phi_-] &= \frac12 \int d^4y \phi_+(y) [\partial^2 - m^2] 
\phi_+(y) - \frac12 \int d^4y \phi_-(y) [\partial^2 - m^2] \phi_-(y)
\nonumber \\
&+ \frac12 \int d^4y \int d^4z \Bigl\{ \phi_+(y) M^2_{++}(y;z) \phi_+(z) +
\phi_+(y) M^2_{+-}(y;z) \phi_-(z) \nonumber \\
& \hspace{1cm} + \phi_-(y) M^2_{-+}(y;z) \phi_+(z) + \phi_-(y) M^2_{--}(y;z) 
\phi_-(z) \Bigr\} \; ,
\end{alignat}
where $M^2_{++}(y;z)$ and $M^2_{--}(y;z)$ are symmetric, and $M^2_{+-}(y;z) 
= M^2_{-+}(z;y)$. The various $\pm$ self-mass-squared functions are related 
by simple rules which alter the $i \epsilon$ terms in propagators and change 
the sign of some vertices \cite{Jordan}. The Schwinger-Keldysh field 
equations are obtained by varying with respect to either $\phi_+$ or 
$\phi_-$, and then equating the two fields after the variation \cite{Jordan},
\begin{equation}
\frac{\delta \Gamma[\phi_+;\phi_-]}{\delta \phi_+(x)} \Bigl\vert_{\phi_{\pm} 
= \phi} \!\!\!\! = [\partial^2 - m^2] \phi(x) - \int d^4x' [M^2_{++}(x;x') + 
M^2_{+-}(x;x')] \phi(x') \; .
\end{equation}
It turns out that $M^2_{+-}(x;x')$ is exactly equal and opposite to 
$M_{++}(x;x')$ whenever $x^{\prime \mu}$ is not in or on the past light-cone
of $x^{\mu}$. When $x^{\prime \mu}$ is in the past of $x^{\mu}$ the two 
terms are complex conjugates, which ensures reality.\footnote{Full details 
can be found in a recent paper on the vacuum polarization of scalar QED in a 
locally de Sitter background \cite{PTW}.}

Since we do not actually derive either the conventional effective action or
the Schwinger-Keldysh version, we will simply employ a trick to extract 
causal field equations from (\ref{genmod}). For the purposes of this paper
one may as well regard these equations --- rather than (\ref{genmod}) ---
as defining the model. The trick is to act the nonlocal operators backwards
whenever they would ordinarily act upon the variation. For example, when
$f[g]$ is any functional of the metric we write,
\begin{eqnarray}
f[g] \frac{\delta \varphi[g]}{\delta g^{\mu\nu}} & = & f[g] \Bigl\{ - 
\frac1{\Box} \frac{\delta \Box}{\delta g^{\mu\nu}} \frac1{\Box} R + 
\frac1{\Box} \frac{\delta R}{\delta g^{\mu\nu}} \Bigr\} \; , \\
& \longrightarrow & \Bigl\{ - \frac{\delta \Box}{\delta g^{\mu\nu}} \varphi
+ \frac{\delta R}{\delta g^{\mu\nu}} \Bigr\} \frac1{\Box} f[g] \; . 
\label{trick}
\end{eqnarray}

It is useful to recall the standard result for varying the Ricci scalar,
\begin{equation}
\frac{\delta R(y)}{\delta g^{\mu\nu}(x)} = \Bigl[ R_{\mu\nu}(y) + D_{\mu} 
D_{\nu} - g_{\mu\nu}(y) \Box \Bigr] \delta^4(y-x) \; . \label{varR}
\end{equation}
(Here $D_{\mu}$ is the covariant derivative operator.) We recall also the
definition of the stress-energy tensor from the variation the matter action
$S_m$,\footnote{This should be local because only gravitons combine the
properties (natural masslessness and the absence of conformal invariance)
needed to produce a strong infrared effect during inflation \cite{TsWo1}.}
\begin{equation}
T_{\mu\nu} \equiv \frac2{\sqrt{-g}} \frac{\delta S_m}{\delta g^{\mu\nu}} \; .
\label{Tmunu}
\end{equation}
Taking $16 \pi G c^{-4}/\sqrt{-g}$ times the variation of our nonlocal
action (\ref{genmod}) --- in the sense of our trick (\ref{trick}) --- gives
the following field equations,
\begin{eqnarray}
\lefteqn{8 \pi G c^{-4} T_{\mu\nu} = 2 [ \Phi_{;\mu\nu} - g_{\mu\nu} \Box \Phi] 
+ G_{\mu \nu} [1 - 2 \Phi]} \nonumber \\
& & + \Bigl[g_{\mu\nu} \varphi^{,\rho} \Phi_{,\rho} - \varphi_{,\mu} 
\Phi_{,\nu} - \varphi_{,\nu} \Phi_{,\mu} \Bigr] + \varphi_{, \mu} \varphi_{, 
\nu} {\cal F}' - \frac{a_0^2}{2 c^4} g_{\mu\nu} {\cal F} \; . \label{MONDeqn}
\end{eqnarray}
The symbol $\Phi[g]$ in this and subsequent formulae denotes the {\it large
potential},
\begin{equation}
\Phi[g] \equiv \frac{1}{\Box} \Bigl( \varphi^{,\rho} {\cal F}' \Bigr)_{;\rho} 
\; . \label{largepot}
\end{equation}

Because we have not really derived (\ref{MONDeqn}) it is worth while to
explicitly verify the important property of conservation. The covariant 
divergence of each term is,
\begin{eqnarray}
2 [ \Phi_{;\mu\nu} - g_{\mu\nu} \Box \Phi]^{;\nu} & = & +2 R_{\mu}^{~\nu}
\Phi_{\! ,\nu} \; . \\
\Bigl( G_{\mu \nu} [1 - 2 \Phi] \Bigr)^{;\nu} & = & -2 R_{\mu}^{~\nu}
\Phi_{\! ,\nu} + R \Phi_{\! ,\mu} \; , \\
\Bigl[g_{\mu\nu} \varphi^{,\rho} \Phi_{,\rho} - \varphi_{,\mu} \Phi_{,\nu} 
- \varphi_{,\nu} \Phi_{,\mu} \Bigr]^{; \nu} & = & - \varphi_{,\mu} \Box \Phi 
- R \Phi_{\! ,\mu} \; , \\
\Bigl(\varphi_{, \mu} \varphi_{, \nu} {\cal F}' \Bigr)^{;\nu} & = & +
\varphi_{; \mu \nu} \varphi^{,\nu} {\cal F}' + \varphi_{,\mu} \Box \Phi \; , \\
\Bigl(- \frac{a_0^2}{2 c^4} g_{\mu\nu} {\cal F} \Bigr)^{;\nu} & = & - 
\varphi_{;\mu \nu} \varphi^{,\nu} {\cal F}' \; ,
\end{eqnarray}
and they do sum to zero. The equations (\ref{MONDeqn}) are manifestly 
covariant. If $1/\Box$ denotes the retarded Green's function they are also
causal.

\section{Spherically symmetric, static sources}

This is the key section of the paper. We begin by working out the small
potential (\ref{smallpot}) and the large potential (\ref{largepot}) for a
general spherically symmetric and static metric (\ref{sphrm}). Then we give
the two independent equations which derive from (\ref{MONDeqn}) for this
geometry. This is valid for arbitrary $A(r) = 1 + a(r)$ and $B(r) = 1 + b(r)$. 
In the MOND regime one has $|a(r)| \ll 1$ and $|b(r)| \ll 1$, for which case
we show that the equations can be simplified without making any assumption
about the interpolating function ${\cal F}(x)$. In terms of the general weak
field ansatz (\ref{ansatz}), these simplified equations prove that $\epsilon_1
+ \epsilon_2 = 0$. It follows from section 2 that no model of the form 
(\ref{genmod}) can be consistent with galaxy lensing. However, it is possible 
to choose the interpolating function ${\cal F}(x)$ so as to reproduce MOND's 
success with galactic rotation curves. The section closes by doing this.

The general spherically symmetric and static geometry (\ref{sphrm},\ref{Gam})
gives rise the following Ricci scalar,
\begin{equation}
R = -\frac{B''}{A B } + \frac{B'}{2 A B} \left(\frac{A'}{A} + \frac{B'}{B}
\right) + \frac{2}{r A} \left(\frac{A'}{A} - \frac{B'}{B}\right) + \frac{2}{
r^2} \left(1 - \frac{1}{A}\right) \; .
\end{equation}
In this geometry, and acting upon a function only of $r$, the covariant 
d'\-Alem\-ber\-tian reduces to,
\begin{equation}
\Box = \frac1{r^2 \sqrt{AB}} \frac{d}{dr} \Bigl( r^2 \sqrt{\frac{B}{A}}
\frac{d}{dr}\Bigr) \; .
\end{equation}
The differential equation which defines the small potential therefore takes
the form,
\begin{equation}
\Bigl( r^2 \sqrt{\frac{B}{A}} \varphi' \Bigr)' = \Bigl( r^2 \sqrt{\frac{B}{A}}
\Bigl[- \frac{B'}{B} + \frac2{r} (A-1) \Bigr] \Bigr)' - r \sqrt{AB} \Bigl(1
- \frac1{A}\Bigr) \Bigl(\frac{A'}{A} + \frac{B'}{B}\Bigr) \; .
\end{equation}
Assuming the two parenthesized terms vanish at $r=0$ we can write,
\begin{equation}
\varphi'(r) = -\frac{B'}{B} + \frac{2}{r} \Bigl(A - 1\Bigr) - \frac{1}{r^2}
\sqrt{\frac{A}{B}} \int_0^r dr' r' \sqrt{AB} \Bigl(1 - \frac1{A}\Bigr) \left(
\frac{A'}{A} + \frac{B'}{B}\right) \; . \label{phi(r)}
\end{equation}

The differential equation that defines the large potential is,
\begin{equation}
\partial_{\mu} \Bigl( \sqrt{-g} g^{\mu\nu} \Phi_{,\nu} \Bigr) = \partial_{\mu}
\Bigl( \sqrt{-g} g^{\mu\nu} \varphi_{,\nu} {\cal F}' \Bigr) \Longrightarrow
\Bigl(r^2 \sqrt{\frac{B}{A}} \Phi'\Bigr)' = \Bigl(r^2 \sqrt{\frac{B}{A}} 
\varphi' {\cal F}' \Bigr)' \; .
\end{equation}
Assuming again that the parenthesized terms vanish at $r=0$ we can write,
\begin{equation}
\Phi'(r) = \varphi'(r) {\cal F}'\!\left(\frac{c^4 \varphi^{\prime 2}(r)}{a_0^2 
A(r)} \right) \; . \label{Phi(r)}
\end{equation}
And those of its second covariant derivatives we shall need are,
\begin{equation}
\Phi_{;tt} = -\frac{B'}{2A} \Phi' \; , \; \Phi_{;rr} = \Phi'' - \frac{A'}{2A}
\Phi' \; , \; \Box \Phi = \frac1{A} \Bigl[\Phi'' + \frac2{r} \Phi' + \frac12
\Bigl( \frac{B'}{B} - \frac{A'}{A} \Bigr)\Bigr] \; .
\end{equation}

In this geometry only the diagonal components of the field equations 
(\ref{MONDeqn}) are nontrivial. The $\theta\theta$ and $\phi\phi$ equations
are proportional to one another, and conservation gives both from the $tt$ 
and $rr$ equations the same as for the stress-energy,
\begin{equation}
\frac{T_{\phi\phi}}{\sin^2(\theta)} = T_{\theta\theta} = \frac{r^3}{2 A}
\Bigl\{ \frac{B'}{2B} \frac{A}{B} T_{tt} + \Bigl[ \frac{d}{dr} + \frac2{r}
- \frac{A'}{A} + \frac{B'}{2B} \Bigr] T_{rr} \Bigr\} \; .
\end{equation}
The system can therefore be defined by its $tt$ (times $A/B$) and $rr$ 
equations,
\begin{alignat}{1}
\frac{8 \pi G A}{c^4 B} T_{tt} &= 2 \Phi'' + \frac{4}{r} \Phi' + \frac{A}{B} 
G_{tt} (1 - 2 \Phi) + \frac{a_0^2}{2 c^4} A {\cal F} - \frac{A'}{A} \Phi' - 
\varphi' \Phi' \; , \\
\frac{8 \pi G}{c^4} T_{rr} &= -\frac{4}{r} \Phi' + G_{rr} (1 - 2 \Phi) - 
\frac{a_0^2}{2 c^4} A {\cal F} - \frac{B'}{B} \Phi' .
\end{alignat}
The $tt$ and $rr$ components of the Einstein tensor are,
\begin{equation}
\frac{A}{B} G_{tt} = \frac{A'}{r A} + \Bigl(\frac{A - 1}{r^2} \Bigr) \qquad ,
\qquad G_{rr} = \frac{B'}{r B} - \Bigl(\frac{A - 1}{r^2}\Bigr) \; .
\end{equation}

We now evaluate the derivative of the small potential (\ref{phi(r)}) to
leading order in the weak fields, $a(r)$ and $b(r)$,
\begin{equation}
\varphi' \longrightarrow \frac{2 a}{r} - b' + \dots \label{smalllim}
\end{equation}
One might worry that the integral in (\ref{phi(r)}) contributes as well, 
but note that the integrand exactly vanishes for $A = B^{-1}$. This means 
that the integral cannot contribute much in the regime for which general 
relativity applies. The integrand is nonzero in the MOND regime, but it 
is also second order in the weak fields, $a$ and $b$. We can therefore 
ignore this term altogether.

In the asymptotic regime we can assume that each derivative adds a factor of
$1/r$. Hence $\varphi'(r)$ goes like $1/r$ times the small numbers $a(r)$ or 
$b(r)$. It follows that $\Phi'/r$ is much larger in magnitude than $\varphi' 
\Phi'$. By similar reasoning we recognize that $\Phi'/r$ and $\Phi''$ dominate 
the other MOND corrections,
\begin{equation}
\Bigl\vert \frac{1}{r} \Phi' \Bigr\vert \sim \Bigl\vert \Phi'' \Bigr\vert
\gg \Bigl\vert \varphi' \Phi' \Bigr\vert \; , \; \Bigl\vert \frac{A'}{A} \Phi'
\Bigr\vert \; , \; \Bigl\vert \frac{B'}{B} \Phi' \Bigr\vert \; , \;
\Bigl\vert \frac{a_0^2}{c^4} {\cal F} \Bigr\vert \; . \label{leading}
\end{equation}

In the weak field limit it seems reasonable to assume $T_{rr} = 0$ while still
allowing a nonzero $A/B \, T_{tt} = \rho$. We still don't know how the two 
leading MOND terms (\ref{leading}) on the gravitational side of the equations
compare with the leading terms from general relativity. Including both gives,
\begin{eqnarray}
2 \Phi'' + \frac{4}{r} \Phi' + \frac{a'}{r} + \frac{a}{r^2} + \dots & = & 
\frac{8\pi G}{c^4} \rho(r) \; \label{total} \\
-\frac{4}{r} \Phi' + \frac{b'}{r} - \frac{a}{r^2} + \dots & = & 0 \; . 
\label{sec}
\end{eqnarray}
The first of these equations (\ref{total}) can be integrated to give,
\begin{equation}
\frac{4}{r} \Phi' + \frac{2 a}{r^2} + \dots = \frac{K}{r^3} + \frac{16 \pi G}{
c^4 r^3} \int_{R_{\rm gal}}^r \!\!\!\!\!\! dr' r^{\prime 2} \rho(r') \; . 
\label{int}
\end{equation}
Adding (\ref{sec}) and (\ref{int}) cancels the leading MOND corrections,
\begin{equation}
\frac{b'}{r} + \frac{a}{r^2} + \dots = \frac{K}{r^3} + \frac{16 \pi G}{c^4 r^3} 
\int_{R_{\rm gal}}^r \!\!\!\!\!\! dr' r^{\prime 2} \rho(r') \; . 
\label{MONDfree}
\end{equation}

Equation (\ref{MONDfree}) is interesting because it has no dependence upon
the still unknown interpolating function ${\cal F}(x)$. We can therefore
use it to make statements about all models of the type (\ref{genmod}). Under 
the assumption of no dark matter halos, the mass integral must eventually 
stop growing, in which case the left hand side falls off like $1/r^3$. So if
$b'(r)$ goes like a constant times $1/r$ then $a(r)$ must go like minus
the same constant. In terms of the generic ansatz (\ref{ansatz}) of section
2 we have just demonstrated that all models of the type (\ref{genmod}) have
$\epsilon_1 + \epsilon_2 = 0$. As discussed in section 2, this means that
galaxies without halos of dark matter would give far too little lensing.
These models do not give phenomenologically viable realizations of MOND.

It is still interesting to see if the interpolating function ${\cal F}(x)$
can be chosen to reproduce MOND rotation curves. For this purpose let us 
consider a sphere of radius $R$ with very low, constant density,
\begin{equation}
\rho(r) = \frac{3 M c^2}{4 \pi R^3} \theta(R - r) \; .
\end{equation}
If the density is small enough the MOND regime prevails throughout, as in a
low surface brightness galaxy. This means that (\ref{total}) can be 
integrated all the way down to $r=0$ to give,
\begin{equation}
2 \Phi' + \frac{a}{r} + \dots = \frac{8 \pi G}{c^4 r^2} \int_0^r \!\! dr' 
r^{\prime 2} \rho(r') \; . \label{firstint}
\end{equation}
We can also use (\ref{smalllim}) to eliminate $b'(r)$ in $-r$ times
(\ref{sec}),
\begin{equation}
4 \Phi' + \varphi' - \frac{a}{r} + \dots = 0 \; . \label{secphi}
\end{equation}
Now eliminate $a(r)$ by adding (\ref{firstint}) and (\ref{secphi}), and then
use (\ref{Phi(r)}) to obtain an equation for the small potential,
\begin{equation}
\varphi' \Bigl[ 1 + 6 {\cal F}'\!\Bigl(\frac{c^4 \varphi^{\prime 2}}{a_0^2}
\Bigr)\Bigr] + \dots = \frac{8 \pi G}{c^4 r^2} \int_0^r \!\! dr' r^{\prime 2}
\rho(r') \; .
\end{equation}

For $r > R$ the mass integral is constant,
\begin{equation}
\varphi' \Bigl[ 1 + 6 {\cal F}'\!\Bigl(\frac{c^4 \varphi^{\prime 2}}{a_0^2}
\Bigr)\Bigr] + \dots = \frac{2 G M}{c^2 r^2} \qquad \forall r > R \; .
\label{penult}
\end{equation}
Now recall from section 2 that MOND requires $\epsilon_2 = 2$, and we have 
just seen that any model of the class (\ref{genmod}) must have $\epsilon_1 = 
-\epsilon_2$. The weak field limit (\ref{smalllim}) for the small potential 
therefore implies we must have,
\begin{equation}
\varphi'(r) \longrightarrow - \frac6{r} \sqrt{\frac{a_0 G M}{c^4}} + \dots
\end{equation}
It follows that the constant term within the square brackets of (\ref{penult})
must exactly cancel, and that the next order term must involve one power of
$\varphi'$. Working out the algebra gives,
\begin{equation}
{\cal F}'(x) = -\frac16 - \frac{\sqrt{x}}{108} + O(x) \Longrightarrow 
{\cal F}(x) = -\frac{x}6 - \frac{x^{\frac32}}{162} + O(x^2) \; .
\end{equation}
The associated weak fields are,
\begin{equation}
a(r) \longrightarrow \frac{4 G M}{3 c^2 r} - 2 \sqrt{\frac{a_0 G M}{c^4}} \; , 
\; b(r) \longrightarrow - \frac{8G M}{3 c^2 r} + 2 \sqrt{\frac{a_0 G M}{c^4}} 
\ln\Bigl(\frac{r}{R}\Bigr) \; . 
\end{equation}
For the general weak field ansatz (\ref{ansatz}) of section 2 we have just
shown $-2 \delta_1 = \delta_2 = -\frac83$ and $-\epsilon_1 = \epsilon_2 = 2$.

A potentially troublesome point is that equation (\ref{penult}) involves 
terms of second order in the weak fields $a(r)$ and $b(r)$,
\begin{equation}
\varphi' \Bigl[ 1 + 6 {\cal F}'\!\Bigl(\frac{c^4 \varphi^{\prime 2}}{a_0^2}
\Bigr)\Bigr] + \dots = \frac{c^2}{18 a_0} \varphi^{\prime 2} + \dots =
\frac{2 G M}{c^2 r^2} \; . \label{ques}
\end{equation}
But in (\ref{leading}) we previously neglected such second order terms to
derive the simplified equations (\ref{total}-\ref{sec}) which pertain in the
MOND regime. Closer inspection reveals that all of the terms neglected in
(\ref{leading}) contribute terms to the left hand side of (\ref{ques}) which
are small for $r \ll R_{\rm hor} \sim 10^{26}~{\rm m}$,
\begin{equation}
r \varphi^{\prime 2} \ll \frac{c^2}{a_0} \varphi^{\prime 2} \sim 10^{27}~{\rm 
m} \times \varphi^{\prime 2} \; .
\end{equation}
Note also that one {\it must} involve quadratic terms like those of 
(\ref{ques}) in order to make the weak fields go like the square root of 
the system's mass, as MOND predicts.

Enforcing the MOND limit determines only the first two terms in the small $x$
expansion of the interpolating function ${\cal F}(x)$. There are many ways of
extending this to a formula for general $x$. The only phenomenological 
constraint is that we need the MOND corrections to be acceptably small in
the general relativistic regime of large $x$. For example, we can make 
${\cal F}(x) \longrightarrow -\frac{14}3 |x|^{\frac12}$ for large $|x|$ with
the following extension,
\begin{eqnarray}
{\cal F}'(x) & = & - \frac{\frac{7}{18} {\rm sgn}(x)}{1 + \frac16 
|x|^{\frac12}} + \frac{\frac{2}{9} {\rm sgn}(x)}{\Bigl(1 + \frac16 
|x|^{\frac12} \Bigr)^2} \; , \\
{\cal F}(x) & = & - \frac{\frac{22}3 |x|^{\frac12} + \frac{7}{9} |x|}{1 + 
\frac16 |x|^{\frac12} } + 44 \ln\Bigl(1 + \frac16 |x|^{\frac12} \Bigr) \; .
\end{eqnarray}
For $|x| \gg 1$ this would typically suppress MOND corrections by some 
characteristic length of the system divided by $c^2/a_0 \sim 10^{27}~{\rm m}$. 
If that is not sufficient one can always extend ${\cal F}(x)$ differently to
obtain more suppression.

\section{Homogeneous and isotropic sources}

Although our relativistic formulation of MOND is ruled out by lensing it 
seems a pity not to work out the cosmology now that we have the formalism. 
The exercise also affords a potentially important caveat on just how much
a general formulation of MOND can change in passing from the static 
geometries of galaxies to the time dependent geometry of cosmology. We begin
by working out the small and large potentials for a homogeneous, isotropic 
and spatially flat metric,
\begin{equation}
ds^2 \equiv -c^2 dt^2 + a^2(t) d\vec{x} \cdot d\vec{x} \; .
\end{equation}
The nonlocal field equations (\ref{MONDeqn}) are next specialized to this
geometry. Then the MOND limit is taken. The section closes by considering
what happens when matter domination follows a long period of radiation
domination.

In this geometry the Ricci scalar is,
\begin{equation}
c^2 R = 6 \dot{H} + 12 H^2 \qquad {\rm where} \qquad H \equiv 
\frac{\dot{a}}{a} \; .
\end{equation}
The small potential is defined by the equation,
\begin{equation}
\Box \varphi(t) = - a^{-3} \frac{d}{dct} \Bigl( a^3 \frac{d\varphi}{dct} 
\Bigr) = R(t) \; .
\end{equation}
If we define the initial values of $\varphi$ and its first derivative to be
zero, it takes the simple form,
\begin{equation}
\varphi(t) = - \int_0^t dt' a^{-3}(t') \int_0^{t'} dt'' a^{3}(t'') \Bigl(
6 \dot{H}(t'') + 12 H^2(t'') \Bigr) \; . \label{phidef}
\end{equation}

The large potential is defined by the differential equation,
\begin{equation}
\partial_{\mu} \Bigl( \sqrt{-g} g^{\mu\nu} \Phi_{,\nu} \Bigr) = \partial_{\mu}
\Bigl( \sqrt{-g} g^{\mu\nu} \varphi_{,\nu} {\cal F}' \Bigr) \Longrightarrow
\frac{d}{dt} \Bigl(a^3 \dot{\Phi} \Bigr) = \frac{d}{dt} \Bigl( a^3
\dot{\varphi} {\cal F}' \Bigr) \; .
\end{equation}
If we again assume null initial values the result is,
\begin{equation}
\Phi(t) = \int_0^t dt' \dot{\varphi}(t') {\cal F}'\!\Bigl(- c^2 a_0^{-2}
\dot{\varphi}^2(t') \Bigr) \; .
\end{equation}
The nonzero components of the second covariant derivative are,
\begin{equation}
\Phi_{;00} = c^{-2} \ddot{\Phi} \qquad , \qquad \Phi_{;ij} = - c^{-2} H
\dot{\Phi} g_{ij} \;.
\end{equation}

We assume the stress-energy tensor to take the perfect fluid form,
\be
T_{\mu\nu} = p g_{\mu\nu} + (p + \rho) u_\mu u_\nu \; .
\ee
Stress-energy conservation implies, $\dot{\rho} = - 3 H (\rho + p)$. The 
nonzero components of the Einstein tensor are,
\begin{alignat}{1}
c^2 G_{00} &= 3 H^2 \; , \\
c^2 G_{ij} &= -(2 \dot{H} +3 H^2) g_{ij} \; .
\end{alignat}
We therefore extract two nontrivial relations from the general equation
(\ref{MONDeqn}),
\begin{alignat}{1}
8 \pi G c^{-2} \rho &= -6 H \dot{\Phi} + 3 H^2 (1 - 2 \Phi) + 
\frac{a_0^2}{2 c^2} {\cal F} \; , \label{rhoeqn} \\
8 \pi G c^{-2} p &= 2 \ddot{\Phi} + 4 H \dot{\Phi} - (2 \dot{H} + 3 H^2 ) 
(1 - 2 \Phi) - \dot{\varphi} \dot{\Phi} - \frac{a_0^2}{2 c^2} {\cal F} \; . 
\label{peqn}
\end{alignat}
As in general relativity, only one of these equations is independent. The
second equation (\ref{peqn}) follows from the first and stress-energy
conservation.

In the MOND regime the interpolating function and its derivative take the
forms,
\begin{equation}
{\cal F}(x) \longrightarrow -\frac1{6} |x| \qquad , \qquad {\cal F}'(x)
\longrightarrow -\frac16 {\rm sgn}(x) \; .
\end{equation}
For cosmology the argument $x = -(c \dot{\varphi}/a_0)^2$ is {\it negative}
so the large potential has the same sign as the small potential,
\begin{equation}
\Phi(t) \longrightarrow \frac16 \varphi(t) \; .
\end{equation}
In the MOND regime we can therefore express (\ref{rhoeqn}) as,
\begin{equation}
-H \dot{\varphi} + 3 H^2 \Bigl(1 - \frac13 \varphi \Bigr) - \frac1{12} 
\dot{\varphi}^2 + \dots = 8 \pi G c^{-2} \rho \; . \label{mondrho}
\end{equation}

An additional specialization of great interest to cosmology is the case of a
power law scale factor,
\begin{equation}
a(t)=\Bigl(1 + H_i t\Bigr)^s \; . \label{power}
\end{equation}
Here $H_i$ is $1/s$ times the Hubble parameter at $t=0$. Substituting into 
(\ref{phidef}) gives the small potential,
\begin{equation}
\varphi(t) = -6 s \Bigl(\frac{2s-1}{3s-1}\Bigr) \Bigl\{\ln\Bigl[1 + H_i 
t\Bigr] - (1-3s)^{-1} \Bigl[\Bigl(1 + H_i t \Bigr)^{1-3s} \!\!\! - 1 \Bigr] 
\Bigr\} . \label{smallphi}
\end{equation}
At late times only the logarithm matters. In this regime we can also express
$\dot{\varphi}$ in terms of the Hubble parameter,
\begin{equation}
\dot{\varphi}(t) \longrightarrow -6 \Bigl(\frac{2s - 1}{3s - 1}\Bigr) H(t) \; .
\end{equation}
We can therefore write the MOND analog of the Friedman equation for power 
law expansion,
\begin{equation}
3 \Bigl\{1 + 2 \sigma - \sigma^2 + 2 s \sigma \ln\Bigl[1 + H_i t\Bigr] \Bigr\}
H^2(t) + \dots = 8 \pi G c^{-2} \rho(t) \; ,
\end{equation}
where $\sigma \equiv (2s-1)/(3s-1)$. The effect of the logarithm on $s > 
\frac12$ power laws is to gradually slow the expansion. This makes rough 
physical sense if we think in terms of MOND strengthening the force of 
gravity in the weak field regime.

For the case of radiation domination ($s=1/2$ and $\sigma=0$) we note that
$\varphi(t) = 0$! The large potential also vanishes --- exactly, not just in 
the MOND regime. Hence the equations reduce to those of general relativity, 
but with the energy and pressure coming from only ordinary matter. This is 
phenomenologically unacceptable. One of many things that goes wrong is 
nucleosynthesis.

\section{Discussion}

We have succeeded in embedding the MOND force law in a set of covariant, 
causal and conserved field equations for the metric. The model suffers from
at least two fatal phenomenological problems in which its predictions agree 
with those of general relativity without dark matter. The deflection of light 
and cosmology during radiation domination both have this property. Of course 
these problems do not necessarily mean that MOND is wrong, only that our 
realization of it is.

Our model consists of corrections which are based on the small potential 
$\varphi[g] = \Box^{-1} R$. One naturally wonders if it is possible to 
find a nonlocal scalar potential that avoids the problems with lensing and
cosmology while still keeping the MOND force law. For example, one might
replace the covariant d'Alembertian with the conformal one,
\begin{equation}
\varphi_c[g] \equiv \frac1{\Box_c} R \qquad {\rm where} \qquad \Box_c \equiv
\Box - \frac16 R \; .
\end{equation}
The distinction between $\Box$ and $\Box_c$ disappears in the weak field
regime because $R$ goes like one power of the weak fields over $r^2$. So such
a model would still give the MOND force law. Unfortunately it would also give
too little lensing precisely because its weak field limit agrees with that
of $\varphi[g]$. Because $\varphi_c$ vanishes with $R$, it would also have
problems dealing with radiation domination without dark matter.

The fact that $R$ vanishes for a radiation dominated universe means that
we should avoid it as the source upon which the nonlocal operator acts. The
next most complicated scalar potential would seem to be,
\begin{equation}
\varphi_2[g] \equiv \frac{c^4}{a_0^2} \frac1{\Box} \Bigl(R^{\mu\nu} 
R_{\mu\nu} \Bigr) \; .
\end{equation}
Because $\varphi_2$ has roughly two derivatives acting upon two powers of the
weak fields, one must also change the Lagrangian,
\begin{equation}
{\cal L}_2 = \frac{c^4}{16 \pi G} \Bigl[ R + c^{-4} a_0^2 {\cal F}_2\Bigl(
\varphi_2[g] \Bigr) \Bigr] \sqrt{-g} \; .
\end{equation}
The interpolating function ${\cal F}_2(x)$ would become linear in the MOND 
regime.

One interesting thing about our model is that it becomes conformally
invariant in the MOND limit. The first hint of this came when all our MOND 
corrections cancelled out of the formula for the deflection of light. Photons
are conformally invariant so they experience no deflection due to conformal
transformations of the metric. 

To prove asymptotic conformal invariance, note that in the MOND limit
(${\cal F}(x) \longrightarrow -\frac16 x$ for $x > 0$) the large potential is,
\begin{equation}
\Phi[g] = \frac{1}{\Box} \Bigl( \varphi^{,\rho} {\cal F}' \Bigr)_{;\rho} 
\longrightarrow -\frac16 \varphi \; .
\end{equation}
In this limit the field equations (\ref{MONDeqn}) take the form,
\begin{equation}
8 \pi G c^{-4} T_{\mu\nu} = \frac13 \Bigl(g_{\mu\nu} \Box \varphi - 
\varphi_{;\mu\nu} \Bigr) + R_{\mu\nu} -\frac12 g_{\mu\nu} R + \dots \; .
\end{equation}
The right hand side is traceless, which means that the linearized theory is
conformally invariant.

This sort of asymptotic conformal invariance was also found by Bekenstein and 
Milgrom \cite{BekMil,Milgrom3}. Conformal invariance was certainly not built 
into either model. For example, the trace of our full field equations 
(\ref{MONDeqn}) is nonzero,
\begin{equation}
8\pi G c^{-4} T^{\mu}_{~\mu} = -6 \Box \Phi - R [1 - 2 \Phi] + 2 \varphi^{,\mu}
\Phi_{,\mu} + \varphi^{,\mu} \varphi_{,\mu} {\cal F}' - \frac{2 a_0^2}{c^4}
{\cal F} \; .
\end{equation}

Asymptotic conformal invariance arises from enforcing the MOND limit. Since 
MOND requires the weak fields to go as the {\it square root} of the source 
mass, it is necessary that terms linear in the weak fields drop out of some 
component of the field equations. The only distinguished component in a 
conserved, tensor equation is the trace. Hence terms linear in the weak fields 
must drop out of the trace of the field equations, which means that the 
linearized theory is conformally invariant.

If asymptotic conformal invariance is generic it means that no metric-based
formulation of MOND can give the required amount of gravitational lensing.
It might also bear on the view that the successes of MOND derive from galaxy 
formation and evolution flowing, through conventional physics, towards some 
sort of fixed point. This is because critical phenomena and universality
are typically characterized by conformal invariance. It might be interesting
if the analogy could be pursued sufficiently to generate quantitative
predictions.

\centerline{\bf Acknowledgments}

It is a pleasure to acknowledge conversations with A. Kosowsky. This work 
was partially supported by DOE contract DE-FG02-97ER\-41029 and by the 
Institute for Fundamental Theory.

\end{document}